\documentstyle[prl,aps,multicol,epsf]{revtex}
\tighten
\begin{document}
\draft

\title{
Influence of 
Quantum Fluctuations on Phase Coherent Andreev Tunneling}
\author{Andrea Huck$^1$, F.W.J. Hekking$^2$, and Bernhard Kramer$^1$}
\address{
$^1$ I. Institut f\"{u}r Theoretische Physik, Jungiusstrasse 9,
D-20355 Hamburg, Germany\\
$^2$Cavendish Laboratory, University of Cambridge,
Madingley Road, Cambridge CB3 OHE, United Kingdom}

\maketitle

\begin{abstract}
We study the subgap transport properties of a small capacitance 
normal metal-superconductor tunnel junction coupled to an
external electromagnetic environment. 
Mesoscopic interference between the electrons in the normal metal
strongly enhances the subgap conductance with decreasing bias voltage. 
On the other hand, quantum fluctuations of the environment 
destroy electronic phase coherence and suppress the subgap conductance 
at low bias (Coulomb blockade).
The competition between charging effects and mesoscopic interference 
leads to a non-monotonic dependence of the differential subgap 
conductance on the applied bias voltage. This feature is pronounced,
even if the coupling to the environment 
is weak and the charging energy is small. 
\end{abstract}


\begin{multicols}{2}

Charge transport through a tunnel barrier between a normal metal (N)
and a superconductor (S) is a widely investigated topic \cite{MT}.
At energies much smaller than the superconducting gap $\Delta$, 
tunneling of single particles is exponentially suppressed. 
Under these conditions,
charge transport through an N-S interface is dominated by
Andreev reflection \cite{AndreevBTK}. If the normal metal and the 
superconductor are separated by a low transparency tunnel barrier,
two-electron tunneling \cite{Wilkins} determines the subgap conductance.
Recently, 
there has been much interest in the subgap properties of mesoscopic
N-S junctions \cite{Hekking94,Beenakker95}. 
The dependence of the subgap conductance of a tunnel junction on 
temperature or applied bias voltage can be very different, depending 
on the precise lay-out of the system under consideration.

If, for instance,  we consider a tunnel junction between a 
superconductor and a 
thin metallic film at low temperatures, electrons move 
phase coherently in N and undergo multiple elastic scattering 
events by impurities or rough sample boundaries.
As a consequence, they will be scattered back to the junction 
interface several times, where they attempt to tunnel into S.
Two-electron tunneling involves two almost time reversed 
electrons. Therefore, the phase of the two-electron tunneling amplitude 
is not randomized by elastic scattering and the amplitudes 
for various tunneling 
attempts add up coherently. This strongly {\em enhances} the subgap conductance
at low bias voltages~\cite{HN,Pothier}. 
On the other hand, in mesoscopic N-S tunnel junctions with 
a small junction capacitance $C$, charging effects\cite{ALIN} become important.
In order to tunnel, the two electrons should overcome the characteristic
Coulomb interaction energy $E_{c}=e^2/C$. This will strongly {\em 
suppress} the subgap conductance~\cite{ex,th} at low temperatures and bias 
voltages $k_{B}T, eV \ll E_{c}$, a phenomenon known as Coulomb blockade of
two-electron tunneling.

\begin{figure}
{\epsfxsize=8.3cm\epsfysize=5cm\epsfbox{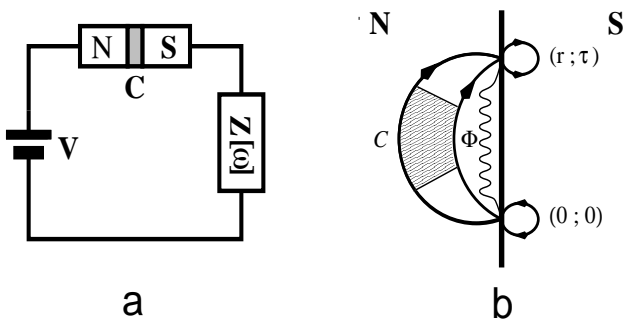}}
\caption{(a) Single N-S junction with a capacitance
$C$ coupled to an external circuit with an impedance $Z(\omega)$ and a 
voltage source. (b) Two electrons propagating 
coherently in a disordered normal metal
as a cooperon ${\cal C}(r;\tau)$ (half-moon) coupled to the
electromagnetic environment by the phase correlator $\Phi(\tau)$
(wavy line). The upper loop describes two electrons
which immediately form a Cooper pair after entering the 
superconductor; the lower loop describes 
the corresponding time-reversed process.}
\label{circcoop}
\end{figure}

In the present paper we will discuss the influence of the competition between 
charging and interference effects on the subgap conductance of a 
single N-S tunnel junction. Charging effects in a single junction are
conveniently described using the so-called electromagnetic environment model
\cite{environment}. In this model, electron tunneling is studied
in the presence of
quantum phase fluctuations due to the Johnson-Nyquist noise of the
external circuit, seen by the junction. 
In the simplest case, this circuit
consists of a capacitor $C$ ({\em i.e.}, the junction capacitance) and an 
external series resistor $Z(\omega) = R$, 
see Fig.~\ref{circcoop}a. 
The influence of such an environment on the subgap conductance
of N-S junctions
has been studied before~\cite{HDBardas97}, but without considering
mesoscopic interference. 
We will show how interference effects are destroyed as 
the series resistance $R$, which determines the coupling to the 
environment, is increased. If the charging energy is smaller than
the superconducting gap,  
the competition between 
quantum fluctuations and mesoscopic interference leads to a
non-monotonic dependence of the differential subgap conductance on 
bias voltage, even for weak coupling.

The system shown in Fig.~\ref{circcoop}a 
can be described by the Hamiltonian $H = H_0 + H_{T}$.
Here, $H_0$ denotes the unperturbed Hamiltonian, 
$H_0=H_N+H_{S}+H_{env}$, where 
$H_{N}$ and $H_{S}$ describe the disordered normal metal and the 
superconductor, respectively.
The electromagnetic environment is described by the usual 
bosonic Caldeira-Leggett\cite{CL} Hamiltonian $H_{env}$.
The tunnel Hamiltonian $H_{T}$ transfers electrons between N and S 
and couples the electrons to the environment; it will be treated 
perturbatively. 
In the interaction picture $H_{T}$ takes the form
\begin{eqnarray}
H_T(t)=\int_N d^3
& r & \int_S d^3r^{\prime}\,\sum_{\sigma}
[
\psi^{\dagger}_{N,\sigma}( r,t)T( r,r^{\prime})
\psi_{S,\sigma}( r^{\prime},t) 
 \nonumber \\
&\times&
e^{-i[eVt/\hbar+\varphi(t)]} + h.c.
]
\,\,.
\end{eqnarray}
Here, $\psi_{i,\sigma}$ is a fermionic field operator for an electron 
with $i = N,S$ and spin $\sigma = \uparrow, \downarrow$;
$T( r,r^{\prime})$ is the amplitude to tunnel from a point $r$ in 
N to $r^{\prime}$ in S, and $V$ the applied 
bias voltage. The phase operator $\varphi(t)$
describes the voltage fluctuations at the tunnel junction, induced
by the electromagnetic environment. Its dynamics is governed by 
$H_{env}$.

From an expansion 
in $H_T$ up to fourth order,
using standard imaginary-time techniques,
one obtains the following expression for the subgap current:
\end{multicols}
\vspace*{-0.2truein} \noindent \hrulefill \hspace*{3.6truein}
\begin{eqnarray}
{\cal I}(\omega_1,\omega_2,\omega_3)=
& &
\frac{24 e t_0^4}{\hbar ^{4}}\,
\mbox{Im}\, 
\int _{B} d^{2}r_{1} d^{2}r_{2} d^{2}r_{3} d^{2}r_{4}
\int_{0}^{\beta \hbar}\,d\tau_{1}d\tau_{2}d\tau_{3}\,
e^{i\omega_{1}\tau_{1}+i\omega_{2}\tau_{2}+i\omega_{3}\tau_{3}} 
\nonumber\\
& & 
\times 
{\cal C}_N(r_{1},r_{2},r_{3},r_{4};0,\tau_{1},\tau_{2},\tau_{3})
\Phi(0,\tau_{1},\tau_{2},\tau_{3}) 
{\cal C}_S(r_{4},r_{3},r_{1},r_{2};\tau_{3},\tau_{2},\tau_{1},0)\, , 
\label{drie}
\end{eqnarray}
\begin{multicols}{2}
\noindent
where $\beta=1/k_BT$ and the 
analytical continuation for the bosonic Matsubara frequencies
$ i\omega_{1}\rightarrow 
-eV/\hbar+i\delta,\,\,
 i\omega_{2}\rightarrow eV/\hbar+i\delta,\,\,
 i\omega_{3}\rightarrow eV/\hbar+i\delta$ has to be performed. 
In order to obtain (\ref{drie}), we assumed tunneling 
to occur between neighboring points on the 
barrier B, located at $z_{B} =0$; correspondingly we put
$T( r,r^{\prime})=t_{0}\delta^3(r-r^{\prime})\delta(z)$.  
The amplitude $t_{0}$ can be expressed in terms of the normal state 
conductance $G_{T}$ of the tunnel barrier, $G_{T} = 
4 \pi t_0^2 S_B N_F/(\hbar v_F R_K)$. Here $R_K = 2 \pi \hbar /e^{2}$ 
is the quantum resistance, $N_F$ the density of states at the Fermi level
and $v_F$ the Fermi velocity; $S_B$ is the area of the barrier surface.
Furthermore, we introduced the four-point correlator
${\cal C}_i(r_{1},r_{2},r_{3},r_{4};\tau_{1},\tau_{2},\tau_{3},\tau_{4})
=
\langle{\hat T}_{\tau}\Psi^{\dagger}_{\uparrow\,i}(r_{1},\tau_{1})
                     \Psi^{\dagger}_{\downarrow\,i}(r_{2},\tau_{2})
                     \Psi_{\downarrow\,i}(r_{3},\tau_{3})
                     \Psi_{\uparrow\,i}(r_{4},\tau_4)
\rangle
$
describing the propagation of two electrons for $i = N,S$, as well as a
four-point phase correlator
$\Phi(0,\tau_{1},\tau_{2},\tau_{3})
=
\langle{\hat T}_{\tau} e^{-i[\varphi(0)+\varphi(\tau_{1})-\varphi(\tau_{2})
-\varphi(\tau_{3})]}\rangle 
$ 
related to the voltage fluctuations.
The averages $\langle \ldots \rangle$ are taken with respect to 
$H_{N}$, $H_{S}$, and $H_{env}$, respectively; in addition the 
correlator ${\cal C}$ is to be averaged over disorder. 

Further simplification can be achieved following Ref.\cite{HN}.
At energies much smaller than the gap $\Delta$, two electrons propagate 
coherently through N over distances 
of the order 
$\xi_{N}=\sqrt{\hbar D/\mbox{max}[eV,k_{B}T,\hbar/\tau_{\varphi}]}$, much 
larger than the corresponding length $\xi_{S}=\sqrt{\hbar D/\Delta}$ in S
($D$ is the diffusion constant and $\tau _{\varphi}$ the phase 
breaking time); moreover, 
the lifetime $\sim \hbar/\Delta$
of a quasiparticle in S in the intermediate state is negligibly small.
Therefore, we have 
${\cal C}_S \sim \delta(r_{1} - r_{2}) \delta(r_{3} - r_{4})
\delta(\tau_{1}) \delta(\tau_{2} - \tau_{3})$ 
in Eq.~(\ref{drie}). 
The dominant contribution to the subgap current can now be written as 
\begin{eqnarray}
I(V)
&=&
\left.\frac{3 R_K G_T^2}{\pi \hbar e S_B N_F^2} \right|
_{i\omega _{\mu}\rightarrow 2eV/\hbar+i\delta}\nonumber \\
& \times &
\mbox{Im}\left[ \int_B\,d^2 r 
      \int_0^{\beta \hbar}\,d\tau e^{i\omega _{\mu}\tau}
      {\cal C}_{N}(r;\tau)
      {\Phi}(\tau)\right]\,\,.
\label{2}
\end{eqnarray}

The integrand of Eq.~(\ref{2}) is depicted diagrammatically in 
Fig.~\ref{circcoop}b. We see
two electrons tunneling from S to N at initial position and time (0;0)
thereby 
interacting with the environment.
The electrons propagate through N to position $r$, where they 
arrive at time $\tau$; their propagation is described by the 
disorder averaged two-particle correlator (Cooperon),
${\cal C}_{N}(r;\tau) =
\langle
{\cal C}_N(0,0,r,r;0,0,\tau ,\tau)
\rangle _{\rm disorder}$.
Then they tunnel back into S, interacting once more with the 
environment.
The wavy line in Fig.~\ref{circcoop}b denotes the phase correlator 
${\Phi}(\tau)
=
\langle{\hat T}_{\tau} e^{-2i[\varphi(\tau)-\varphi(0)]}\rangle
=
\exp{[4J(\tau)]}$, where we choose 
$J$ according to the electromagnetic environment model ~\cite{environment}:
\begin{eqnarray}
J(\tau)
&=&
-2\,\int^{\infty}_0\,\frac{d\omega}{\omega} 
\frac{\mbox{Re}\, 
 Z_t(\omega)}{R_K} \nonumber \\
& \times &
\left(\coth(\beta\frac{\hbar \omega}{2})
(1-\cosh(\omega \tau)) + \sinh(\omega |\tau|)\right).
\label{J}
\end{eqnarray}
Here $Z_t(\omega) = 1/(i\omega C + 1/Z(\omega))$ is
 the total impedance seen by 
the junction.

Upon performing the analytic continuation in Eq.~(\ref{2}) 
the pair tunneling current finally is found to be
\end{multicols}
\vspace*{-0.2truein} \noindent \hrulefill \hspace*{3.6truein}
\begin{eqnarray}
I(V)=\frac{3 R_K G_T^2 }{2 \pi^2 e S_B N_{F}}  
\int_{B} d^2r
\int^{\infty}_{-\infty}
dE dE^{\prime}\,C_{N}(r;E) P(E^{\prime})
\nonumber\\
\times
\frac{1-\exp[-2eV\beta]}{1-\exp[(E^{\prime}-2eV)\beta]}
 (f[(E-2eV+E^{\prime})/2]-f[(E+2eV-E^{\prime})/2])
 .
\label{5}
\end{eqnarray}
\begin{multicols}{2} \noindent
The function 
$C_{N}( r;E)$ is the spatial fourier transform of the real part of
the diffusion propagator $1/[-iE+\hbar DQ^2+\hbar/\tau_{\varphi}]$.
The probability $P(E)$ to emit or absorb a photon with 
frequency $E/\hbar$ during tunneling is defined as
\begin{equation}
P(E)
=
\frac{1}{2\pi \hbar}
\int
dt
\exp{[4J(t) +i E t/\hbar]} ,
\label{FT}
\end{equation} 
where $J(t)$ can be obtained from Eq.~(\ref{J}) by putting 
$\tau =it$ for $\tau \ge 0$.

As an example, we will study the Andreev current (\ref{5}) of a N-S 
tunnel junction,
consisting of a quasi one-dimensional (1D)
normal metal wire in contact with a superconductor via a tunnel barrier
with dimensions much smaller then $\xi_N$. 
For a quasi 1D wire with cross section $S_W$, we have
$$
\int _{B} d^{2}r C_N(r,E) = \frac{S_{B}}{S_{W}} 
\frac{\cos[\arctan(E \tau_{\varphi}/\hbar)/2]}
{2 \sqrt{\hbar D}(E^{2}+(\hbar/\tau_{\varphi})^{2})^{1/4}}
$$ 
in Eq.~(\ref{5}).
The junction has a capacitance $C$ and is
embedded in a purely resistive environment 
$Z(\omega)=R$. The total impedance is thus given by
$Z_{t}(\omega)=1/(i\omega C+R^{-1})$.
However, the calculation of $P(E)$ can only be performed numerically  
in this case~\cite{Falci91}. Therefore,
in order to proceed analytically, we will use the
approximation $\mbox{Re} [Z_{t}(\omega)]= R \exp(-\omega/\omega_{c})$,
which has the correct zero-frequency limit $\mbox{Re} [Z_{t}(0)]= R$.
The cut-off frequency is chosen to be $\omega_c=E_c R_K /\hbar R$, such that  
the approximated impedance also gives the correct 
phase correlation function $J(t)\sim \exp(-i2 E_c t/\hbar)$ 
in the limit of short times. In particular, this guarantees that this 
approximation yields
the correct behavior for $R=\infty$, namely the 
complete suppression of tunneling for $eV<E_c=e^2/C$.

At zero temperature the phase correlation function is readily found to be
$\Phi(t)=1/[1+i\omega_c t]^{\alpha}$; its fourier transform
(\ref{FT}) yields the probability distribution
\begin{equation}
P(E)=\frac{e^{-E/\hbar \omega_c}}{\Gamma(\alpha)\hbar\omega_c} 
\left(\frac{E}{\hbar\omega_c}\right)^{\alpha-1}\theta(E) .
\label{PF1}
\end{equation}
The negative part of the spectrum is truncated, since at $T=0$
photons can only be emitted.
The power $\alpha=8R/R_{K}$ can be interpreted as a parameter 
determining the coupling strength between the 
electronic phase and the environment. 

\begin{figure}
{\epsfxsize=7cm\epsfysize=5.7cm\epsfbox{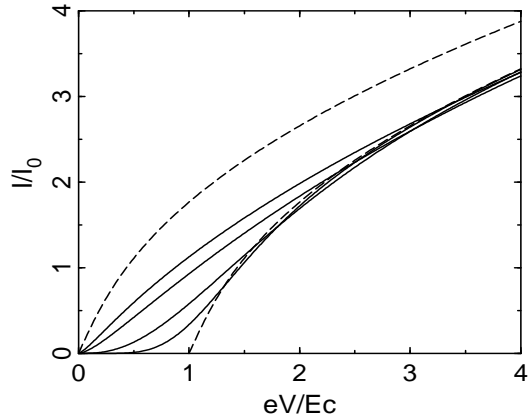}}
\caption{Andreev current in units of
$I_0 =(3 G_T^2/2\pi\sigma S_W  e)$ $\times \protect\sqrt{\hbar D E_c}$
 for $T=0$ and $\hbar/\tau_{\varphi}=0.5 E_c$.
Curves from top to bottom correspond to $\alpha=0$ (dashed line),
$\alpha=1/4,\, 1/2,\, 2,\, 8$ (solid lines) 
and $\alpha=\infty$ (dashed line).}
\label{cu}
\end{figure}

For the non-interacting system, 
$\alpha =0$, the subgap conductance is proportional to the
''coherence resistance'' $R_{coh}=\xi _{N}/\sigma S_{W}$ \cite{HN}, where
$\sigma$ denotes the conductivity of the wire.
In the limit of strong coupling, $\alpha=\infty$, a gap
appears in the I-V curve below $E_c$.
We will focus on the case of small charging energies $E_c\ll \Delta$.
For a finite coupling and
finite  phase coherence time $\tau_{\varphi}$~\cite{footnote}, two regimes exist
(see Fig.~\ref{cu}): 
(i)
For very small voltages $eV\ll\hbar /\tau_{\varphi}$,
 interference is cut off by 
$\tau_{\varphi}$ and the I-V curve shows a power law
behavior $I\propto V^{(\alpha+1)}$.
This is what one  would expect for
noninterfering electrons. 
(ii) For higher voltages $eV\gg\hbar /\tau_{\varphi}$
 the coherence length is voltage dependent
and the I-V characteristic changes: $I \propto V^{\alpha+1/2}$.
Remarkably, if  
$\tau_{\varphi}=\infty$ and $\alpha=\alpha_c=1/2$, suppression  
of the current by charging effects and  
enhancement by interference cancel each other exactly
 at voltages below $E_c/e$, such that the
 I-V curve is linear.
For values of $\alpha$ larger than the ''critical'' coupling $\alpha_c$,
the power is always larger than one and a Coulomb gap starts to evolve
with increasing $\alpha$. 

\begin{figure}
{\epsfxsize=7cm\epsfysize=5.7cm\epsfbox{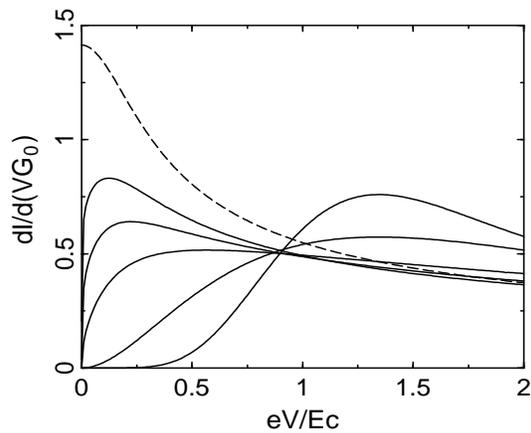}}
\caption{Differential conductance in units of 
$G_0=$ $(3 G_T^2 /2 \pi \sigma S_{W})\protect\sqrt{(\hbar D/ E_c)}$ 
for $T=0$ and $\hbar/\tau_{\varphi}=0.5 E_c$.
The maximum evolves to the right as $\alpha$ is increased from 
$\alpha=0$ (dashed line), taking
$\alpha=1/8,\, 1/4,\, 1/2,\, 2,\,8$ (solid lines).}
\label{dc}
\end{figure}

The differential conductance $G(V)=dI/dV$ is strongly suppressed at 
$eV < \hbar/\tau_{\varphi}$ for {\em arbitrary } $\alpha>0$. 
The zero bias peak, which is the 
fingerprint of phase coherent Andreev tunneling, is destroyed by charging 
effects. 
Instead, the differential conductance will display a peak at finite bias 
(Fig.~\ref{dc}).
Increasing the voltage on the one hand lifts
the Coulomb blockade, but on the other hand decreases the coherence length. 
If $\alpha <\alpha _c$, the maximum
in the differential conductance
appears at a bias
$eV \alt\hbar /\tau_{\varphi}$ and will shift to zero bias 
for $\tau_{\varphi}=\infty$. The coupling to the quantum fluctuations 
of the environment is too weak
to fully destroy phase coherence. Therefore coherent pair tunneling is 
blocked 
only at voltages below $\hbar /\tau_{\varphi}$, where interference is cut off.
For $\alpha > \alpha_c$, the coupling is strong enough for charging effects
to dominate the behavior. The conductance is suppressed in the entire
voltage regime below $E_c$ by quantum fluctuations, 
which makes $\tau_{\varphi}$ superfluous as
a cutoff for the divergence of $G$ at zero bias. 
The maximum appears at $eV \agt E_c$ and shifts towards $E_c$ 
as $\alpha$ is increased.

At finite temperatures, the electrons 
can gain energy by absorbing environmental modes. 
Therefore, 
the Coulomb blockade is gradually lifted 
with increasing temperature. 
The probability distribution $P(E)$ at small energies 
$|E| \ll \hbar \omega_c$ 
can be calculated analytically, following~\cite{Ingold94}. 
In the long time limit, the phase correlator reads
$\Phi(t)\simeq[(\pi k_BT/\hbar \omega_c)/ \sinh(\pi t k_BT/\hbar)]^{\alpha}$. 
From (\ref{FT}) we find
\begin{equation}
P(E)\simeq\frac{\exp(E/2k_BT)}{2\pi \Gamma(\alpha)\omega_c}
\left[\frac{2 \pi k_B T}{\hbar \omega_c}\right]^{\alpha-1}\left| 
\Gamma \left(\frac{\alpha}{2}+\frac{i E}{k_B T}\right) \right|^2 \,.
\label{PF2}  
\end{equation}
For low temperatures $k_{B}T \ll \hbar \omega_{c}$ one easily calculates 
the zero bias 
differential conductance 
$G(T,V=0)$ with the help of (\ref{PF2}). 
For $k_B T\ll \hbar/\tau_{\varphi}$ we find
$G(T,V=0)\sim T^{\alpha}$,
whereas $G(T,V=0)\sim T^{\alpha -1/2}$ for
$k_B T\gg \hbar/\tau_{\varphi}$.

\begin{figure}
{\epsfxsize=7cm\epsfysize=5.7cm\epsfbox{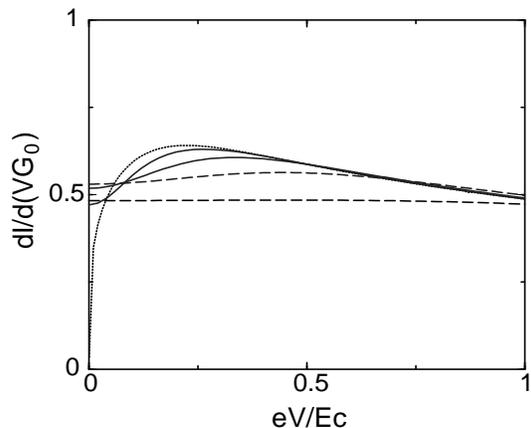}}
\caption{Sketch of differential conductance for $\alpha < \alpha_c$.
From top to bottom, temperature increases: $T=0$ (dotted line),
$k_B T \ll \hbar/\tau_{\varphi}$ (solid lines), and $k_B T \protect\alt 
\hbar/\tau_{\varphi}$ (dashed lines).}  
\label{dct}
\end{figure}

In order to determine $G(T,V)$ away from zero bias, the function $P(E)$
should be calculated for arbitrary $E$. For nonzero temperatures, this can only
be done numerically. Qualitatively, one expects the Coulomb blockade 
to be lifted if $k_{B}T \agt \hbar/\tau_{\varphi}$
for weak coupling, $\alpha < \alpha _{c}$, whereas for $\alpha > \alpha _{c}$ 
thermal smearing becomes relevant at temperatures $k_{B}T \agt E_c$.
As an example, the 
differential conductance as a function of bias voltage is sketched in
Fig.~\ref{dct} for $\alpha < \alpha _{c}$ at various temperatures .

{\bf Acknowledgments}. The authors would like to thank 
D. Averin, R. Fazio, 
A. van Otterlo, and
M. Sassetti for useful discussions. 
The financial support of the European Union
(Contract 
ERB-CHBI-CT941764) is gratefully acknowledged.
Part of this work was done at the ISI Torino and the ICTP Trieste.

\narrowtext

\end{multicols}


\begin{references}
\bibitem{footnote} In typical experiments, like
Ref.~\cite{Pothier}, $\hbar/\tau_{\varphi}$ is relatively large 
($\sim$ 10 mK) 
at the lowest temperatures.
\end{references}
\end{document}